# A computational hierarchy in human cortex

*Andreea O. Diaconescu[1,2], Vladimir Litvak[3], Christoph Mathys[1,4,5], Lars Kasper[1,6], Karl J. Friston[3], Klaas E. Stephan[1,3]*

[1]*Translational Neuromodeling Unit, Institute for Biomedical Engineering, University of Zürich & ETH Zürich, Switzerland;* [2]*University of Basel, Department of Psychiatry (UPK), Basel, Switzerland;* [3]*Wellcome Trust Centre for Neuroimaging, University College London, London, UK;* [4]*Scuola Internazionale Superiore di Studi Avanzati (SISSA), Trieste, Italy;* [5]*Max Planck UCL Centre for Computational Psychiatry and Ageing Research, London, UK;* [6]*Institute for Biomedical Engineering, MRI Technology Group, ETH Zürich & University of Zurich, Switzerland;*

Correspondence should be addressed to:

Andreea O. Diaconescu, PhD
Translational Neuromodeling Unit (TNU)
Institute for Biomedical Engineering,
University of Zürich and ETH Zürich
Wilfriedstrasse 6  CH-8032 Zürich
diaconescu@biomed.ee.ethz.ch
+41 44 634 91 09

# Abstract

Hierarchies feature prominently in anatomical accounts of cortical organisation. An open question is which computational (algorithmic) processes are implemented by these hierarchies. One renowned hypothesis is that cortical hierarchies implement a model of the world's causal structure and serve to infer environmental states from sensory inputs. This view, which casts perception as hierarchical Bayesian inference, has become a highly influential concept in both basic and clinical neuroscience. So far, however, a direct correspondence between the predicted order of hierarchical Bayesian computations and the sequence of evoked neuronal activity has not been demonstrated. Here, we present evidence for this correspondence from neuroimaging and electrophysiological data in healthy volunteers. Trial-wise sequences of hierarchical computations were inferred from participants' behaviour during a social learning task that required multi-level inference about intentions. We found that the temporal sequence of neuronal activity matched the order of computations as predicted by the theory. These findings provide strong evidence for the operation of hierarchical Bayesian inference in human cortex. Furthermore, our approach offers a novel strategy for the combined computational-physiological phenotyping of patients with disorders of perception, such as schizophrenia or autism.

The notion of hierarchy is central to neurobiological accounts of brain organisation. Anatomical tract tracing studies have revealed a hierarchical organisation of cortical areas, based on identifying ascending (bottom-up) and descending (top-down) connections with specific laminar patterns (*1–3*). Remarkably consistent cortical hierarchies can be derived from laminar patterns of cytoarchitecture (*4*). These structural hierarchies have classically been interpreted as the basis of sensory processing streams and of the associated variations in spatial (*5*) and temporal (*6–8*) receptive fields across processing levels.

By contrast, the actual computations executed by cortical hierarchies are highly debated. One leading proposal derives from predictive coding (*9, 10*) and related theories that view the cortex as an organ for hierarchical Bayesian inference (*11–13*). This theory suggests that cortical hierarchies embody an internal ("generative") model of the world that recapitulates the causal structure of the environment. Such a model would enable probabilistic predictions about how environmental states cause noisy sensory inputs. Conversely, inverting this generative model would allow for hierarchical Bayesian inference on the state of the world; this corresponds to perception.

The results of various recent experimental and theoretical studies on human perception are consistent with the idea of hierarchical Bayesian inference as an algorithmic principle of human cortex (e.g., (*14–18*)). However, central predictions of the hierarchical Bayesian account of brain function remain empirically untested. Most importantly, a direct correspondence between the sequence of hierarchical computations as predicted by the theory and the empirically observed sequence of computation-specific neuronal responses has not been demonstrated yet. In this paper, we present evidence for this correspondence using multimodal imaging and a cognitive task requiring multi-level learning.

We obtained functional magnetic resonance imaging (fMRI, N=47) and electroencephalography (EEG, 128 channels, N=48) data from healthy volunteers. The participants performed a variation of an established social learning task (*19*); for details, see (*20*). In this deception-free task, participants predicted the trial-wise outcome of a binary lottery, with veridical probabilities displayed as a pie chart (55-75%; Figure 1).

We refer to this pie chart information as "cue". Additional advice was provided by a videotaped adviser who had more accurate (80%) information but also incentives to switch between helpful or misleading advice as the task proceeded. Participants were informed truthfully that advisers had their own (non-disclosed) incentives and that their intention to help or mislead might change over time. In order to optimally integrate the advice with the cue (pie chart), participants thus not only had to infer (i) on the accuracy of current advice, but also (ii) on the intention behind it and (iii) on how this intention might change in time (volatility). In other words, on each trial, the participants faced a hierarchical inference problem with three levels.

A hierarchical inference process of this sort can be parsed into a sequence of belief updates that, under generic assumptions, are governed by two quantities: prediction errors (PEs) and precision weights (*20, 21*). The latter are crucially important since they determine how strongly PEs drive belief updates (*21*). We inferred subject-specific expressions of trial-wise belief updating sequences, including the underlying PEs and precision weights, from the participants' expressed behaviour. For this, we employed the hierarchical Gaussian filter (HGF (*21*)), a commonly used model for computational analyses of behaviour in terms of hierarchical Bayesian inference (e.g., (*20, 22–24*)). Notably, we performed an initial model selection procedure to (i) verify that the HGF provided a better explanation of the participant's behaviour than other common models, and to (ii) determine the most likely belief-action mapping (response model); see Methods for details.

Importantly, the HGF specifies a concrete order in which PEs and precision weights must be computed in order to update beliefs across all levels of the hierarchy (Figure 2A). This allowed us to test for a direct correspondence between the predicted sequence of computations and the temporal order of associated brain responses. To this end, we conducted trial-by-trial analyses of EEG and fMRI data, using a general linear model (GLM) that was informed by the subject- and trial-specific estimates of PEs and precision weights (for details, see Methods). In brief, at the single-subject level, the regression model contained all computational regressors of interest shown by Figure 2A; these regressors were not orthogonalised with respect to each other. For each computational quantity, we thus estimated its unique contribution to explaining brain activity in voxel space (fMRI) and at all EEG sensors over a peri-stimulus time window of [50:550ms] (relative to outcome onset). Our analysis focused on the belief updating process (at the sharply defined time points of trial outcome), and not on the predictions (whose exact timings during the video-clip based advice delivery was uncertain). Random effects group analysis across all participants was performed using a standard summary statistic approach.

The results of the sensor-level EEG analysis are summarised by Figure 3. We used F-tests to identify significant brain responses and report effects that survived whole-brain family-wise error (FWE) correction at the cluster-level ($p < 0.05$), with a cluster-defining threshold (CDT) of $p<0.001$ that ensures valid inference (see Methods). It can be seen that the temporal order of activity associated with the different computational variables precisely matches the hierarchical processing sequence as prescribed by the HGF (compare Figure 2A). From a cognitive perspective, this mirrors the hierarchical form of inference in our task, from non-social to social quantities and with increasing degrees of abstraction.

Beginning at the bottom of the computational hierarchy, the three low-level PEs (with respect to cue, advice, and outcome) and belief precision about the advice occured first, as predicted by the model. The PEs were associated with EEG activity peaks at 134 ms, 166 ms and 258 ms at occipital, posterior, and occipital-temporal channels, respectively. Interestingly, activity associated with the advice PE differed depending on whether the advice was perceived as helpful or misleading: it peaked at 166 ms in posterior channels for positive PE trials (when advice was more helpful than predicted) and at 168 ms in frontal channels for negative PE trials (advice was more misleading than predicted). Belief precision about the advice followed with a positive peak at 352 ms over posterior-central channels.

As predicted by the model, PEs that updated the volatility of the adviser's intentions came next, showing a positive activity peak at 398 ms in frontocentral channels. This was followed by the associated precision weight at 534 ms over posterior-central channels – an order again in line with the predictions by our model.

In summary, this sensor-level EEG analysis found a clear match between the predicted order of hierarchical computations within each trial of the social learning task and the temporal sequence of associated EEG responses. Notably, this correspondence between the sequence of evoked responses and the model's predictions is unlikely to arise by chance (p = 0.033; see Methods).

One might ask why the activations do not seem to be spatially aligned in a perfectly posterior-anterior fashion, as one might expect for an underlying anatomical hierarchy. Such an expectation, however, should be taken with a grain of salt, for two reasons: First, anatomical hierarchies are defined by connectional and cytoarchitectonic criteria (*1–4*) but do not strictly follow a posterior-anterior gradient (*25*); for example, in the visual system, the frontal eye field is situated at a low level of the hierarchy (*3*). Second, the analysis reported above was conducted at the sensor level; this makes it difficult to link the evoked potentials to specific cortical regions and does not reveal the sources of activity related to the different computational quantities. To address the second issue, we used the fMRI data from the same task as spatial priors to guide EEG source reconstruction and test whether the observed correspondence between sequences of computational steps and neuronal activations would also hold in source space.

All of the computational quantities implied by our hierarchical Bayesian model also gave significant results in trial-by-trial fMRI analyses, surviving whole-brain correction at the cluster-level (p<0.05; with a CDT of p<0.001; Figure 2B, Table S3). These fMRI activations by trial-wise PEs and precison weights mapped onto classical "theory of mind" regions; in particular, precisions and PEs about advice accuracy and the adviser's intentions were localised in cortical areas involved in mentalising functions, such as the middle cingulate gyrus, medial prefrontal cortex and temporo-parietal junction (*26–28*). Using these fMRI results as spatial priors for constrained source localisation (Figure 4), we identified the cortical sources of the EEG activity shown in Figure 3 (for details of the source reconstruction procedure, see Methods).

For each significant peak detected in the sensor-level ERP analysis (red labels in Figure 4), we found a temporally matching counterpart at the level of cortical source activity (grey bars in Figure 4). These cortical sources explained, on average, 94% of the signal variance within the time window of the ERP effects. Starting at the bottom of the hierarchy, the activity peak elicited by cue PEs at 134 ms (after trial outcome) was localised to the right lingual gyrus. This was followed by the advice PE (right anterior temporo-parietal junction at 166 ms), and the outcome PE activity (at 258 ms in the right superior occipital cortex). Subsequently, the activity peak of advice precision at 352 ms was localised to right superior frontal gyrus, the volatility PE activity peak at 398 ms to right dorsal anterior cingulate gyrus, and finally, the volatility precision activity peak at 534 ms to right dorsal middle cingulate gyrus (Figure 4).

In conclusion, sensor-level EEG and fMRI-guided source space EEG analyses consistently demonstrated a direct match between the sequence of computations prescribed by the HGF as a generic model of hierarchical Bayesian inference and the temporal order of cortical activations elicited by these computations. This provides empirical support for hierarchical Bayesian inference as a central algorithmic principle of cortex, as proposed by predictive coding and related "Bayesian brain" theories (*9–13*). Our analyses illustrate how a combination of multimodal imaging and computational modelling of behaviour can track the operation of cortical algorithms in space and time. This approach may prove useful for clinical purposes. Given that aberrant hierarchical Bayesian inference has been implicated in the pathophysiology of schizophrenia (*29, 30*) and autism (*24*), sensitive probes are required that can detect subtle disturbances of hierarchical inference in the temporal domain. The computationally informed single-trial EEG analysis presented in this paper may usefully complement traditional model-based fMRI methods and finesse the computational phenotyping of patients with mental disorders.

## Materials and Methods

### Participants

95 healthy male right-handed volunteers (EEG: N=48; fMRI: N=47) between 19 and 30 years participated in two studies (age: 23±0.49 in the EEG study and 24±0.43 in the fMRI study [mean±SE]). Both samples corresponded to placebo groups from pharmacological studies whose results will be reported in future publications. There was no overlap between the participants recruited for fMRI and EEG, as we wanted them to be naïve with respect to the social learning task. Participants gave written informed consent before entering the study. Ethics approval was obtained by the Ethics Committee of the Canton of Zurich (KEK-ZH-Nr. 2012-0567).

Participants had normal or corrected-to-normal vision and were healthy as indicated by medical history and clinical examination including electrocardiography prior to participation. Smokers or any individuals with a previous history of neurological or psychiatric diseases or drug abuse were excluded from participation.

### Experimental Design

#### Stimuli and social learning task

The stimuli were selected from a previous behavioural study using face-to-face interactions(*31*). In this initial study, participants were randomly assigned to a "player" or an "adviser" role. The player predicted the outcome of a binary lottery with probabilities displayed in the form of a pie chart (varying from 55-75%). Players accrued points with every correct prediction; their final payment was proportional to the total score plus a potential bonus if the score exceeded pre-defined silver or gold targets (*Figure 1A, top panel*). The adviser's role was to instruct the player which option to choose by holding up either a green or a blue card. The adviser based his suggestion on information he received (the probability of this information being valid was always 80%). Furthermore, the adviser could monitor the player's progress with respect to his own silver or gold targets; importantly, these targets differed from those of the player and provided an incentive structure that motivated the adviser to alternate between giving helpful or misleading advice throughout the game (*Figure 1A, lower panel*). Players were truthfully informed that advisers had distinct incentives which could motivate them to change their intentions during different phases of the game. Additionally, they were told that advisers received more accurate, albeit incomplete information; thus, the advice could also be unintentionally correct or incorrect.

Based on the main strategy employed by the advisers, two of the recorded full-length videos were edited into 2 second segments for the purpose of the present study. All video clips were matched in terms of their luminance, contrast, and colour balance using the video editing software Adobe Photoshop Premiere CS6. One of the two chosen advisers was randomly assigned to each participant and no differences in performance and degree of reliance on the advice were observed between the two adviser groups: t(47)=-0.5652, *p*=0.57 and t(46)=0.2327, *p*=0.81, respectively.

## Procedure

The social learning task described above was adapted for the present studies as follows: Participants were presented with the binary lottery and video clips of the advisers indicating their recommendations for 2 seconds. Subsequently, they were asked to predict the outcome (and indicate the prediction by button press) during a 5 second decision phase that was followed by the presentation of the outcome, i.e., the winning colour (*Figure 1b*). In total, the experiment consisted of 210 trials which contained 6 visual cue types (75:25, 65:35, 55:45, 45:55, 35:65, and 25:75 % blue: % green pie charts).

## MRI data acquisition

MRI data were acquired using a Philips Ingenia 3T whole-body scanner with a 32-channel SENSE head coil (Philips Medical Systems, Best, The Netherlands) at the Institute for Biomedical Engineering, University of Zurich and ETH Zurich. In both the fMRI and EEG studies, high-resolution inversion-recovery T1-weighted 3D-TFE (turbo field echo) structural images were acquired for each participant (301 slices; voxel size 1.1×1.1×0.6 mm$^3$; FOV 250 mm; TE 3.4 ms).

In the fMRI study, we additionally acquired gradient echo T2*-weighted echo-planar images (EPIs) with blood-oxygen-level dependent (BOLD) contrast (33 slices/volume; TR 2.5 s; voxel size 2×2×3 mm$^3$; interslice gap 0.6 mm; FOV 192×192×180 mm; TE 36 ms; flip angle 90°). Oblique-transverse slices with +15° right-left angulation were acquired. The experimental task was run in two sessions to give participants a short break to move. There were 740 and 580 volumes in the first and the second session, respectively, together with five discarded volumes at the start of each session to ensure T1 effects were at equilibrium.

Stimuli were projected onto a display (NordicNeuroLab LCD MR-compatible 32-inch monitor) which participants viewed through a mirror fitted on top of the head coil. Participants' heart rate and respiration was recorded with a 4-electrode electrocardiogram and a breathing belt.

The functional images were realigned and co-registered to the participant's own structural scan using SPM12 (http://www.fil.ion.ucl.ac.uk/spm/; version 6906). The structural image was processed using a unified segmentation procedure combining segmentation, bias correction, and spatial normalisation (32); the same normalisation parameters were then used to normalise the EPI images. Finally, EPI images were smoothed with a Gaussian kernel of 6mm full-width half-maximum.

Correction for physiological noise was performed with the PhysIO toolbox (33) (http://www.translationalneuromodeling.org/tapas) using Fourier expansions of different order for the estimated phases of cardiac pulsation (3rd order), respiration (4th order) and cardio-respiratory interactions (1st order).

### EEG Data acquisition and preprocessing

EEG was recorded at a sampling rate of 500 Hz using a BrainVision system with 128 scalp electrodes. The horizontal and vertical electrooculogram (EOG) was recorded from channels attached infraorbitally and supraorbitally to the left eye.

Pre-processing and data analysis was performed using SPM12. Continuous EEG recordings were referenced to the average, high-pass filtered using a Butterworth filter with cut-off frequency 0.5 Hz and low-pass filtered using Butterworth filter with cut-off frequency 48 Hz. We focused our analyses on the amplitude modulations following the presentation of the outcome. The data were epoched into -500 to 550 ms segments around the presentation of the outcome. An EEG forward model (3-shell Boundary Element Model (34)) was specified based upon (i) individual electrode locations digitally recorded during the experiment and (ii) canonical head meshes (inverse) normalized to match the subjects' structural MRI scan (35).

We removed eye-blink related artefacts by applying a multiple source eye correction method (36), as implemented in SPM12, version 6906. Eye blink events were identified with a thresholding approach applied to the vertical EOG data; these events were used to epoch the continuous EEG into 1000 ms segments (-500 ms to 500 ms around these events). Ocular source components were determined using singular value decomposition (SVD) of topographies from all the trials and all the time points and combined with a set of cortical topographies generated by SVD of the cortical mesh lead fields. The pseudo-inverse of the resulting matrix was then used to remove the spatial subspace spanned by the eye-blink components, but orthogonal to the brain components from the data epoched around the outcome presentations. We used three ocular components. This number was determined empirically by starting at one and increasing the number until there was no further reduction in the amplitude of the average eye-blink when corrected with the same method.

Finally, an additional artifact rejection procedure was applied using a thresholding approach to detect problematic trials or channels. Trials in which the signal recorded at any of the channels exceeded 100 μV relative to the pre-stimulus baseline were removed from subsequent analysis. We also visually inspected the trials to verify that these outliers were artefactual. The total number of artefactual trials that were rejected was 149, with an average of 3.17 ± 4.3 trials per participant and a range of 0-19 trials across subjects.

## Strategy for Model-Based Analysis of Single-Trial EEG data

Following preprocessing and artefact correction as described above, single-trial EEG data were converted to 3-dimensional volumes with two spatial dimensions (anterior-to-posterior and left-to-right directions of the scalp surface) and a temporal dimension (peri-stimulus time). These scalp × time 3D images can be subjected to statistical analysis using the general linear model (GLM) in an analogous fashion to fMRI(37). This is a well-established approach for statistical analysis of scalp EEG data using the SPM software(38). For both imaging modalities, we pursued single-trial analyses where a computational model of behavioural responses provided trial-wise predictions.

In the following, we describe the computational models we considered and how trial-wise predictions from the chosen model entered GLMs of voxel time-series and evoked EEG responses, respectively.

## Model Space

We formalised our hypotheses about the mechanisms underlying the players' observed behaviour in terms of a hierarchical model space with 12 models including both Bayesian and reinforcement learning models(39), resulting in a 3×2×2 factorial model space (for details, see (20, 31) and Figure S1). Model parameters were estimated using a gradient-based optimization method (the BFGS algorithm(40)) and compared using random effects Bayesian model selection (BMS(41)).

## Perceptual Model: Hierarchical Gaussian Filter

The hierarchical Gaussian filter(21, 42) (HGF) is a hierarchical Bayesian model that captures subject-specific approximations to ideal hierarchical Bayesian inference. It comprises a hierarchy of hidden states $x_1^{(k)}, x_2^{(k)}, \ldots, x_n^{(k)}$ that cause the sensory inputs the agent experiences on each trial $k$. These states evolve in time as Gaussian random walks where the step-size is controlled by the state of the next-higher level. The highest level is assumed to evolve with a constant step-size, and the lowest level gives rise to the experimental stimuli the agent encounters.

In our case, $x_1$ represents a categorical variable, i.e., the advice accuracy; any single piece of advice is either accurate ($x_1^{(k)} = 1$) or inaccurate ($x_1^{(k)} = 0$). All states higher than $x_1$ are continuous. State $x_2$ represents the adviser's fidelity in logit space. The highest state $x_3$ represents the phasic log-volatility of the advisers' intentions. The exact equations describing these relations and the overall generative model are described elsewhere(42).

The HGF parameters capture the individual learning style of participants and determine the evolution of their beliefs in time. Here, we estimated parameters $\kappa$, $\omega$, and $\vartheta$ that denote the strength of the coupling between the second and third level, the tonic log-volatility on the second level, and the variability of volatility over time (meta-volatility), respectively.

**Inversion of the Model: The update equations**

The HGF update equations are derived by variational model inversion and provide approximately Bayes-optimal rules for trial-by-trial updating of an agent's beliefs, given this agent's particular set of parameter values(42). "Belief" refers to a posterior probability distribution as described by its sufficient statistics.

On trial $k$, an observed trial outcome $u$, which indicates that advice was either accurate ($u^{(k)} = 1$) or inaccurate ($u^{(k)} = 0$), leads to a hierarchical cascade of belief updates described by the update equations. At the bottom level, there is complete correspondence between observation $u^{(k)}$, posterior belief $\mu_1^{(k)}$, and state $x_1^{(k)}$ because the accuracy of the advice is seen by the participant without ambiguity:

$$\mu_1^{(k)} = u^{(k)} = x_1^{(k)} \qquad (1)$$

However, the observed outcome $u^{(k)}$ induces a prediction error (PE) $\delta_1^{(k)}$ with respect to the prediction $\hat{\mu}_1^{(k)}$ (the agent's belief about the probability of the advice being correct after the previous trial; see Eq. 5):

$$\delta_1^{(k)} = u^{(k)} - \hat{\mu}_1^{(k)} \qquad (2)$$

The ensuing precision-weighted PE updates are hierarchically organized in the sense that the agent needs to use $\delta_1^{(k)}$ to update its second-level belief about the fidelity $x_2^{(k)}$ of the adviser. $x_2^{(k)}$ is continuous, assumed to be Gaussian and thus represented by the sufficient statistics $\mu_2^{(k)}$ (mean) and $\pi_2^{(k)}$ (precision, i.e., inverse variance). The update $\Delta\mu_2^{(k)} = \mu_2^{(k)} - \hat{\mu}_2^{(k)}$ to the prediction $\hat{\mu}_2^{(k)} = \mu_2^{(k-1)}$ is driven by $\delta_1$ and weighted by $\pi_2$ (42) :

$$\Delta\mu_2^{(k)} = \frac{1}{\pi_2^{(k)}} \delta_1^{(k)} \qquad (3)$$

This update in turn leads to a PE, $\delta_2^{(k)}$, induced by $\Delta\mu_2^{(k)}$. At the third level, the agent's belief about the phasic log-volatility $x_3^{(k)}$ of the adviser's fidelity is represented by the sufficient statistics $\mu_3^{(k)}$ and $\pi_3^{(k)}$, and the pattern from the second level repeats. Specifically, the update $\Delta\mu_3^{(k)} = \mu_3^{(k)} - \hat{\mu}_3^{(k)}$ to the prediction $\hat{\mu}_3^{(k)} = \mu_3^{(k-1)}$ is driven by $\delta_2$ and weighted by $\pi_3$ (42):

$$\Delta\mu_3^{(k)} \propto \frac{1}{\pi_3^{(k)}} \delta_2^{(k)} \qquad (4)$$

After performing these belief updates about the adviser's fidelity and the volatility of his fidelity, the agent is able to update the probability $\hat{\mu}_1^{(k+1)}$ that the advice on the next trial will be correct. This corresponds to the logistic sigmoid of the current expectation of adviser fidelity:

$$\hat{\mu}_1^{(k+1)} = s\left(\mu_2^{(k)}\right) = \frac{1}{1 + \exp\left(-\mu_2^{(k)}\right)} \qquad (5)$$

With every new trial, another cycle of these hierarchically cascading updates takes place. Since the updates are precision-weighted PEs, our analysis includes the PEs that drive the updates ($\delta_1$ and $\delta_2$) and the precisions that weight them ($\pi_2$ and $\pi_3$).

However, PEs relating to the accuracy of the advice ("advice PE") are not the only ones involved in the task. The simplest PE relates to the outcome $u^{(k)}$ relative to the cue $\tilde{c}^{(k)}$ ("cue PE"):

$$\delta_{\tilde{c}}^{(k)} = u^{(k)} - \tilde{c}^{(k)}. \qquad (6)$$

Another, more complicated PE relates to the outcome $u^{(k)}$ relative to predicted outcome $\mu_b^{(k)}$ ("outcome PE"):

$$\delta_b^{(k)} = u^{(k)} - \mu_b^{(k)}, \qquad (7)$$

where the predicted outcome is the weighted average of the predictions $\hat{\mu}_1^{(k)}$ from advice and $\tilde{c}^{(k)}$ from the cue:

$$\mu_b^{(k)} = \zeta\, \hat{\mu}_1^{(k)} + (1-\zeta)\tilde{c}^{(k)}, \qquad (8)$$

where $\zeta$ is the individually estimated social (advice) weight parameter of each participant.

This results in a hierarchy of six PEs and precisions that inform beliefs and their updates in the winning model of our task. In order, they are

1. Cue PE $\delta_{\tilde{c}}$
2. Advice PE $\delta_1$
3. Outcome PE $\delta_b$
4. Precision of belief about advice fidelity $\pi_2$
5. Volatility PE $\delta_2$
6. Precision of belief about volatility $\pi_3$

While the first four quantities do not depend on each other computationally, they are all low-level PEs and precisions. Here, they are ordered by computational and conceptual complexity as reflected by the equations for $\delta_{\tilde{c}}$, $\delta_1$, and $\delta_b$ above. Advice belief precision is given by

$$\pi_2^{(k)} = \frac{1}{1/\pi_2^{(k-1)} + \exp\left(\kappa \mu_3^{(k-1)} + \omega\right)} + \hat{\mu}_1^{(k)}\left(1 - \hat{\mu}_1^{(k)}\right). \qquad (9)$$

The HGF model of our task predicts that these three low-level PEs and the advice belief precision are computed first, before the following high-level quantities can be computed. According to the model, this should occur in the following, strictly defined order:

$\delta_2$ depends on $\pi_2$ directly and on $\delta_1$ because it contains $\Delta\mu_2$:

$$\delta_2^{(k)} = \frac{1/\pi_2^{(k)} + \left(\Delta\mu_2^{(k)}\right)^2}{1/\pi_2^{(k-1)} + \exp\left(\kappa\mu_3^{(k-1)} + \omega\right)} - 1, \qquad (10)$$

$\pi_3$ depends on $\delta_2$:

$$\pi_3^{(k)} = \frac{1}{1/\pi_3^{(k-1)} + \vartheta} + \frac{\kappa^2}{2} w_2^{(k)}\left(w_2^{(k)} + r_2^{(k)}\delta_2^{(k)}\right), \qquad (11)$$

where we have used the notation and definitions of (*42*).

This creates a hierarchical computational architecture in which PEs and precisions are successively elaborated and passed to higher levels, thereby predicting the temporal order in which the associated neurophysiological events should occur. A combinatorial analysis shows that, under this hierarchy, there are 4! sequences that are consistent with the prescribed temporal order (with the three low-level PEs and advice precision in any order, but volatility PE and volatility precision assigned to fifth and sixth computation step), out of a total of 6! possible sequences.

## Response Models

Response models map the agent's beliefs onto decisions(*43, 44*). As participants had access to both the advice and the binary lottery, we modeled their beliefs about outcomes as the integration of the two sources of information (see Equations 7 and 8).

Responses were modelled using a softmax rule, in which the decision temperature was modulated by the perceived volatility of the adviser's intentions, as in(*20, 31*). Responses were coded as $y = 1$ for taking the advice, $y = 0$ for rejecting it:

$$p\left(y^{(k)} = 1 \middle| \mu_b^{(k)}\right) = \frac{\mu_b^{(k)\beta}}{\mu_b^{(k)\beta} + (1 - \mu_b^{(k)})^\beta}, \qquad (14)$$

where $\beta$ represents the inverse decision temperature.

Using the same set of priors for the model parameters as in the initial study (*31*) , maximum-a-posteriori (MAP) estimates of model parameters were obtained using the HGF toolbox version 3.0 (http://www.translationalneuromodeling.org/tapas).

We used family-level inference(*45*) to determine (i) the most likely class of perceptual models, combining across all response models, and (ii) the most likely class of response models, combining across all perceptual models.

Model comparison reproduced previous findings (*20*, *31*), showing that the three-level HGF outperformed competing perceptual models in explaining choice behaviour in the two studies (Tables S1 and S2) and that participants used both (social) advice and (non-social) cues to predict the outcome and infer the adviser's current intentions.

## General Linear Model

Following Bayesian model comparison, we extracted the trajectories of the computational quantities from the winning model and entered them into a subject-specific design matrix. To identify fMRI or EEG correlates of PEs and precisions, we used a model with the computational variables as explanatory variables. This GLM was used to explain either voxel time-series (fMRI study) or observed ERP responses at the single trial level, over channels and peristimulus time (EEG study).

For the fMRI study, the following regressors (plus their temporal and dispersion derivatives) were included in the model (*Figure 2A*); these were event-related regressors of the outcome presentation, parametrically modulated by the respective computational quantity:

1. Outcome x Cue PE ($\delta_{\tilde{c}}$ in Equation 6);
2. Outcome x Advice PE ($\delta_1$ in Equation 2);
3. Outcome x Outcome PE ($\delta_b$ in Equation 7);
4. Outcome x Advice precision ($\pi_2$ in Equation 9).
5. Outcome x Volatility PE ($\delta_2$ in Equation 10).
6. Outcome x Volatility Precision ($\pi_3$ in Equation 11).

The same subject-specific design matrix was constructed for the EEG study with all events time-locked to the outcome presentation.

At the second (group) level, we used a standard summary statistic approach(46) to test the null hypothesis that the first-level parameter estimate of interest was zero across subjects, at any given voxel (fMRI) or at any given sensor and peristimulus time point (EEG). This produced a group-level SPM of the F-statistic of PE or precision effects. We used Gaussian random field theory (47) to perform whole-brain family-wise error (FWE) correction at the cluster-level ($p<0.05$) under a cluster-defining threshold (CDT) of $p<0.001$ that ensures valid inference(48, 49). For EEG, the different SPMs were then summarised using a combined maximum intensity projection map of significant activations, at each point in peristimulus time and over posterior-to-anterior and left-to-right scalp locations, respectively (*Figure 3*).

Furthermore, to link the single-trial EEG results at the sensor level to our fMRI results and determine the cortical sources of the different computational quantities, we applied multiple sparse priors (MSP) source reconstruction (23) to the grand-averaged parameter estimate trajectories ($\hat{\beta}$) over within-trial time, obtained from solving the ordinary least squares (OLS) equation associated with our sensor-level GLM. Each estimated $\hat{\beta}$ waveform reflected the unique contribution of each regressor (computational variable) in explaining the data while factoring out the effects of the other regressors contained in the design matrix. Thus, we obtained six $\hat{\beta}$ waveforms at each electrode and within-trial time point for each subject, reflecting the contribution of each computational variable included in the design matrix. Since these waveforms were produced by a linear transformation of the original trials, they could be subjected to conventional source reconstruction. Source reconstruction was applied to the grand-average waveforms while restricting the sources with a mask computed from the corresponding fMRI contrast, so that only sources appearing in fMRI were allowed to explain the ERP data. The fMRI masks were binary and included all clusters of voxels that survived whole-brain FWE correction at the cluster-level ($p < 0.05$) under a CDT of $p < 0.001$ at the voxel level based on GRF theory (22). Out of the sources representing the computational variables in voxel space (*Figure 2B*), we isolated the region explaining the largest percent variance in the EEG signal at the peak identified by the single-trial EEG analysis (*Figure 3*).

## Author Contributions:

AOD, CM and KES conceived and designed the experiments; AOD performed the experiments and acquired the data; AOD, VL and LK analysed the data; VL designed the EEG analysis pipeline; LK designed the fMRI data acquisition sequence; AOD, VL, CM, LK, KJF and KES wrote the paper.

## Ethics:

All participants gave written informed consent before entering the study. Ethics approval was obtained by the Ethics Committee of the Canton of Zurich (KEK-ZH-Nr. 2012-0567).

# Figures and Legends:

## Figure 1
**Social Learning Experimental Paradigm:**

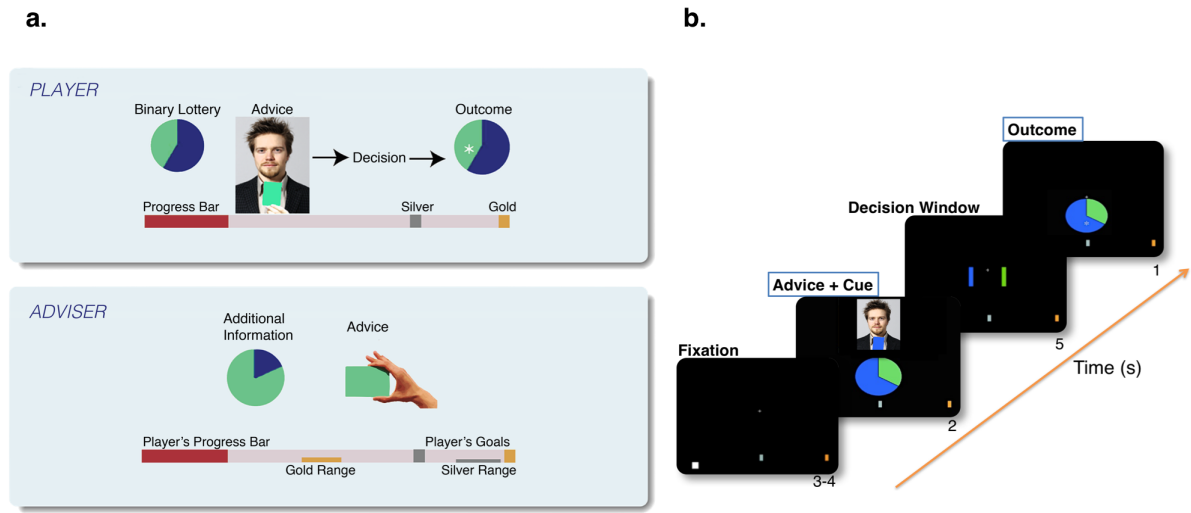

**Figure 1**

(A) Healthy volunteers (fMRI: N=47; EEG: N=48; all male) took part in an advice-taking game for monetary rewards. "Players" predicted the outcome of a binary lottery whereas "Advisers" advised Players on which option to choose. Each Player interacted with one (randomly assigned and pre-recorded from a previous interactive version of the task) Adviser during the entire game. Both roles were incentivized with monetary rewards, and the incentive structure differed to ensure the presence of different learning phases, including both collaboration and competition between the two participants. For the Players, the incentive structure remained stable across time. Players benefited from the Adviser's recommendations as Advisers always received more information about the outcome of the lottery (constant probability of 80%). However, the Advisers' motivation to provide valid or misleading information varied during the game as a function of their own incentive structure. Importantly, Players were (truthfully) informed that the adviser had his own undisclosed incentives and thus his intentions could change during the game. (B) Within-trial timeline of this social learning and inference task during neuroimaging (fMRI and EEG). Video clips of the Advisers were presented for 2 seconds.

### Figure 2
**Computational hierarchy and Associated fMRI Activations:**

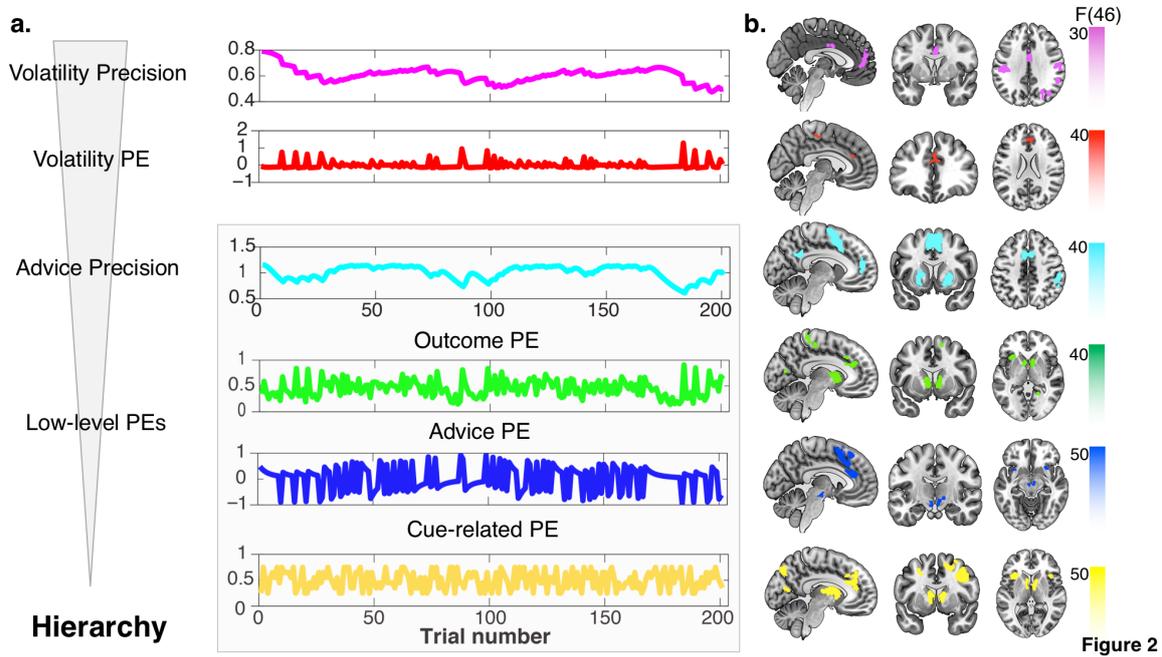

(A) The HGF model of the present task predicts that three types of (low-level) PEs and precison are computed first (with no specified order amongst them): (i) the cue-related PE, the difference between the actual outcome and the outcome predicted by the cue (pie chart); (ii) the advice PE, the discrepancy between the actual accuracy of the advice (correct or misleading) and the participant's expectation of its accuracy; (iii) the outcome PE, the difference between actual trial outcome and the outcome as predicted on the basis of the expected advice accuracy and the cue; (iv) advice precision (the participant's confidence about the fidelity of the adviser). The computational quantities that the HGF model predicts to follow are (v) the volatility PE (which serves to update the estimated volatility of the adviser's fidelity), and (vi) volatility precision. See Supplementary Material for mathematical details.

(B) In fMRI analyses, we used a GLM that included regressors encoding the subject-specific trial-wise PEs and precisions as predicted by the HGF model. The figure shows the results of F-tests that report the significance of the different PEs and precisions under whole-brain FWE cluster-level correction ($p < 0.05$) with a cluster-defining threshold of $p < 0.001$. Starting from the bottom of the hierarchy, (i) the cue-related PE activated the right lingual gyrus, left SPL, right anterior insula, bilateral caudate, and medial and right dorsolateral PFC; (ii) the advice

PE activated the bilateral VTA, right anterior TPJ, left superior parietal lobule, bilateral anterior insula, ACC, and dorsomedial PFC; (iii) the outcome PE activated the bilateral superior occipital gyrus, striatum, ACC, left anterior insula, and bilateral medial PFC; (iv) the advice precision activated the bilateral ACC, posterior cingulate cortex, putamen, and superior frontal gyrus/SMA; (v) the volatility PE activated the bilateral superior frontal gyrus and dorsal ACC, and (vi) the volatility precision activated the bilateral middle cingulate gyrus, ventromedial PFC, right posterior STS and right precentral gyrus.

# Figure 3
**Temporal Evolution of the Computational Hierarchy from EEG:**

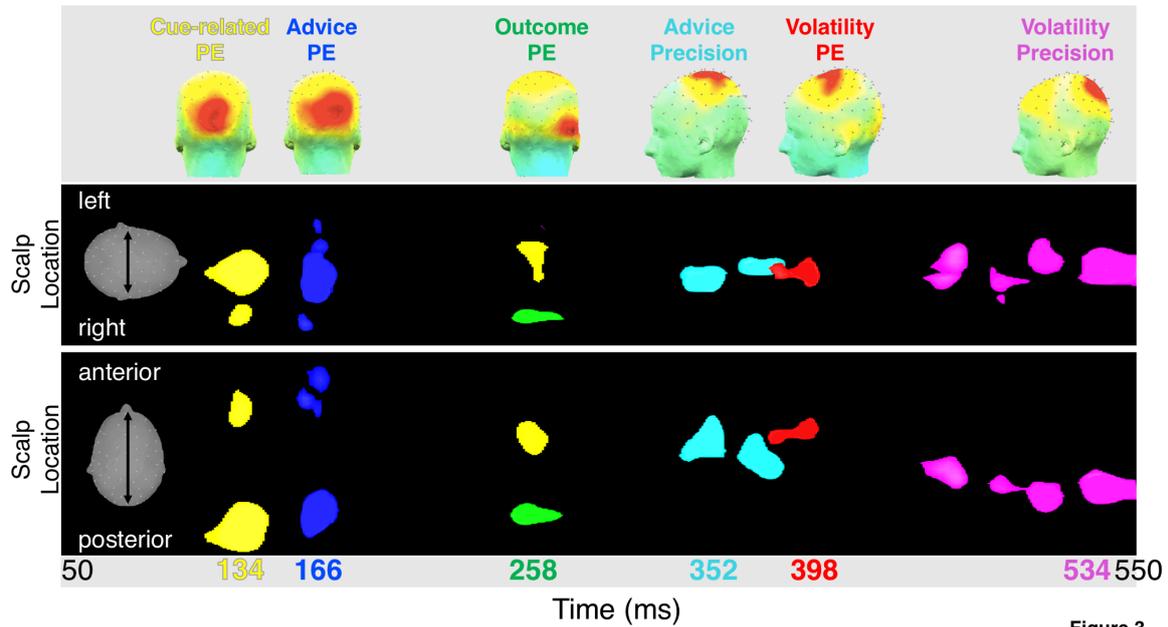

For the EEG analyses, we used a GLM that included regressors representing the subject-specific trial-wise PEs and precisions as predicted by the HGF model. This GLM was used to explain observed event-related potential (ERP) responses at the single trial level, over channels and peristimulus time. We used F-tests to obtain a statistical parametric maps (SPMs) that represented the effect of each PE and precision quantity on evoked responses over all channels and peristimulus time points. These SPMs were thresholded using whole-volume FWE cluster-level correction ($p < 0.05$) with a cluster-defining threshold of $p < 0.001$. This procedure was performed separately for each PE and precision quantity, and the significant effects were combined in a single (color-coded) figure that shows the maximum intensity projections (MIPs) over posterior to anterior and right to left channels, respectively. Time is with respect to outcome onset. The temporal order of activity evoked by the different computational quantities precisely matches the sequence of hierarchical Bayesian inference processes as prescribed by the HGF; compare Figure 2A.

# Figure 4
**Source Extraction Using fMRI Spatial Priors:**

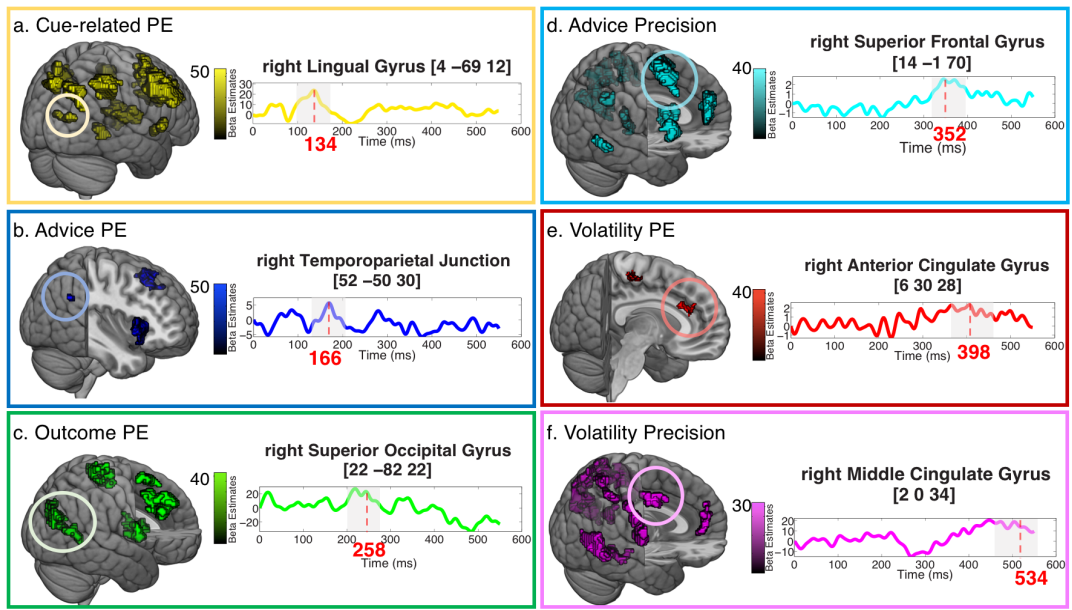

**Figure 4**

Multiple sparse priors (MSP) source reconstruction and source extraction of the computational representations based on spatial priors from fMRI study (see Figure 2). Time is with respect to outcome onset. The significant time-points of the model-based EEG results at the sensor level (see Figure 3) are highlighted in gray and the peak effects for each computational quantity are marked in red.

# Supplementary Material:

## Figure S1:

**Hierarchical structure of the model space**: perceptual models, response models, specific models:

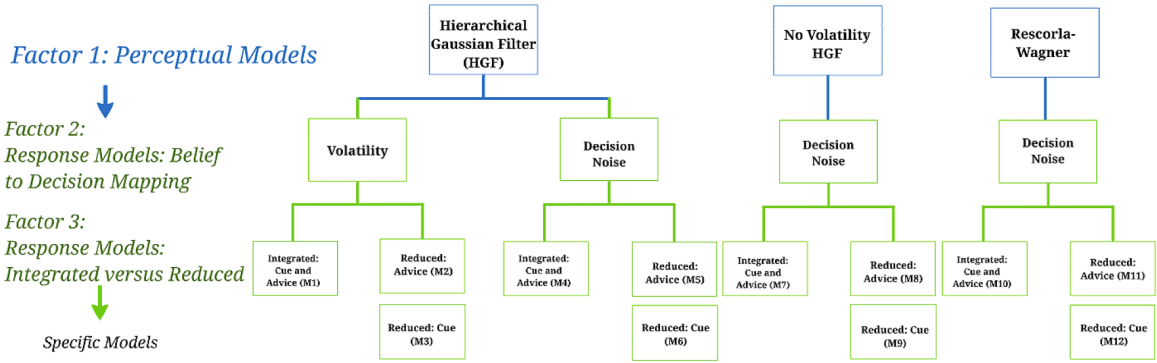

**Figure S1**

The models considered in this study have a 3 x 2 x 2 factorial structure. The specific models at the bottom represent individual models of social learning in which both social and non-social sources of information are considered. The nodes at the highest level represent the perceptual model families (three-level HGF, reduced no-volatility HGF and RW). Two response models were formalized under the HGF model: decision noise in the mapping of beliefs to decisions either (1) depended dynamically on the estimated volatility of the adviser's intentions ('Volatility' model) or (2) was a free parameter over trials ('Decision noise' model). At the second level, the response model parameters can be divided further according to the weighing of social and non-social information—these models assume that participants' predictions are based on (1) both cue and advice information and (2) advice, or (3) cue probabilities (pie chart) only. [reprinted from Diaconescu et al., 2014].

# Table S1

## A. Results of Bayesian model selection (EEG Study): protected exceedance probabilities (xp)

|              | HGF Volatility | HGF Decision Noise | No volatility HGF | RW     |
|--------------|----------------|--------------------|-------------------|--------|
| **Cue & Advice** | 0.9306     | 0.002              | 0.002             | 0.0576 |
| **Advice**   | 0.0052         | 0.002              | 0.0003            | 0.0003 |
| **Cue**      | 0              | 0                  | 0                 | 0      |

## B. Family-level inference (EEG Study: perceptual model set): Posterior model probability or p(r|y) and exceedance probabilities (xp)

|        | HGF    | No Volatility HGF | Rescorla-Wagner |
|--------|--------|-------------------|-----------------|
| **p(r|y)** | 0.9599 | 0.0249            | 0.0152          |
| xp     | 1      | 0                 | 0               |

## C. Family-level inference (EEG Study: family model set): Posterior model probability or p(r|y) and exceedance probabilities (xp)

|        | Integrated | Reduced: Advice | Reduced: Cue |
|--------|------------|-----------------|--------------|
| **p(r|y)** | 0.9857 | 0.0123          | 0.0020       |
| xp     | 1          | 0               | 0            |

# Table S2

## A. Results of Bayesian model selection (fMRI Study): protected exceedance probabilities (xp)

|  | HGF Volatility | HGF Decision Noise | No volatility HGF | RW |
|---|---|---|---|---|
| **Cue & Advice** | 0.9361 | 0.0409 | 0.0001 | 0.0002 |
| **Advice** | 0.02 | 0.0027 | 0 | 0 |
| **Cue** | 0 | 0 | 0 | 0 |

## B. Family-level inference (fMRI Study: perceptual model set): Posterior model probability or p(r|y) and exceedance probabilities (xp)

|  | HGF | No Volatility HGF | Rescorla-Wagner |
|---|---|---|---|
| **p(r|y)** | 0.8818 | 0.0299 | 0.0883 |
| xp | 1 | 0 | 0 |

## C. Family-level inference (fMRI Study: family model set): Posterior model probability or p(r|y) and exceedance probabilities (xp)

|  | Integrated | Reduced: Advice | Reduced: Cue |
|---|---|---|---|
| **p(r|y)** | 0.8482 | 0.15 | 0.0018 |
| xp | 1 | 0 | 0 |

## Table S3

Montreal Neurological Institute (MNI) coordinates and F-values of maxima of fMRI activations by the 6 key computational quantities which were significant under whole-brain family-wise error (FWE) correction at the cluster-level (p<0.05; with a cluster-defining threshold of p<0.001). Related to Figure 2.

|  | Hemisphere | x | y | z | *F* |
|---|---|---|---|---|---|
| **1. Cue-related PE** |  |  |  |  |  |
| superior medial PFC | R | 6 | 36 | 36 | 53.46 |
| anterior insula | L | -30 | 20 | -4 | 39.63 |
| dorsolateral PFC | R | 42 | 6 | 46 | 31.76 |
| caudate | L | -8 | 8 | 10 | 30.72 |
| superior parietal lobule | L | -32 | -48 | 48 | 26.94 |
| anterior insula | R | 28 | 16 | -8 | 26.68 |
| angular gyrus | R | 40 | -50 | 46 | 25.79 |
| superior parietal lobule | R | 34 | -42 | 46 | 24.48 |
| precuneus | R | 14 | -78 | 46 | 24.05 |
| caudate | R | 10 | 8 | 10 | 22.95 |
| lingual gyrus | R | 4 | -69 | 12 | 22.16 |
| lingual gyrus | L | -6 | 72 | 12 | 19.35 |

|  | Hemisphere | x | y | z | *F* |
|---|---|---|---|---|---|
| **2. Advice PE** |  |  |  |  |  |
| anterior insula | L | -38 | 18 | -4 | 73.48 |
| anterior insula | R | 34 | 18 | -2 | 52.34 |
| dorsomedial PFC | R | 0 | 16 | 54 | 43.24 |
| superior parietal lobule | L | -32 | -46 | 42 | 41.87 |
| dorsolateral PFC | R | 48 | 18 | 4 | 41.81 |
| anterior cingulate cortex | R | 8 | 30 | 26 | 36.04 |
| temporo-parietal junction | R | 52 | -50 | 30 | 31.9 |
| superior frontal cortex | R | 20 | 52 | 30 | 26.65 |
| caudate | L | -8 | 2 | 10 | 26.08 |
| ventral tegmental area/substantia nigra | L | -2 | -20 | -16 | 21.58 |
| ventral tegmental area/substantia nigra | R | 4 | -16 | -10 | 20.99 |

|  | Hemisphere | x | y | z | *F* |
|---|---|---|---|---|---|
| **3. Outcome PE** |  |  |  |  |  |
| caudate | L | -8 | 8 | 2 | 31.45 |
| superior medial frontal gyrus | R | 6 | 40 | 36 | 30.23 |
| cuneus | L | -2 | -82 | 26 | 30.16 |
| anterior cingulate gyrus | R | 6 | 34 | 24 | 29.52 |
| superior occipital gyrus | R | 22 | -82 | 22 | 26.9 |
| anterior cingulate gyrus | L | -8 | 34 | 24 | 26.73 |
| SMA | R | 6 | 22 | 58 | 24.13 |

| | Hemisphere | x | y | z | *F* |
|---|---|---|---|---|---|
| superior frontal gyrus | R | 4 | -18 | 56 | 23.77 |

| | Hemisphere | x | y | z | *F* |
|---|---|---|---|---|---|
| **4. Advice Precision** | | | | | |
| supplementary motor area | R | 6 | 8 | 50 | 39.73 |
| supramarginal gyrus | R | 62 | -20 | 26 | 36.72 |
| temporal parietal junction | R | 46 | -52 | 28 | 35.06 |
| supplementary motor area | L | -4 | 12 | 54 | 34.66 |
| putamen | R | 22 | 12 | -4 | 31.72 |
| medial prefrontal cortex | R | 0 | 50 | 24 | 31.07 |
| putamen | L | -18 | 14 | -4 | 30.07 |
| posterior cingulate cortex | L | -8 | -46 | 28 | 20.22 |

| | Hemisphere | x | y | z | *F* |
|---|---|---|---|---|---|
| **5. Volatility PE** | | | | | |
| anterior cingulate cortex | R | 6 | 30 | 28 | 43.30 |
| anterior cingulate cortex | L | -4 | 36 | 22 | 34.66 |
| superior frontal gyrus | R | 4 | -22 | 54 | 19.14 |

| | Hemisphere | x | y | z | *F* |
|---|---|---|---|---|---|
| **6. Volatility Precision** | | | | | |
| superior temporal sulcus | L | -52 | -62 | 14 | 76.46 |
| insula | L | -40 | -2 | 8 | 34.51 |
| middle temporal gyrus | R | 52 | -62 | 8 | 32.69 |
| middle temporal gyrus | R | -54 | -58 | -2 | 31.17 |
| parahippocampal cortex | L | -28 | -44 | -10 | 28.34 |
| angular gyrus | R | 44 | -62 | 18 | 27.53 |
| middle cingulate cortex | R | 4 | -2 | 42 | 23.83 |
| middle cingulate cortex | R | 2 | 0 | 34 | 20.58 |
| temporal parietal junction | R | 56 | -44 | 34 | 20.47 |
| supramarginal gyrus | L | -48 | -32 | 54 | 20.47 |
| ventromedial PFC | L | -4 | 46 | 0 | 20.15 |

## Acknowledgements:

We acknowledge support by the UZH Forschungskredit (AOD), SNF Ambizione PZ00P3_167952 (AOD), the René and Susanne Braginsky Foundation (KES), the Joint Initiative involving Max Planck Society and University College London on Computational Psychiatry and Aging Research (CM), NCCR Neural Plasticity and Repair (KES, LK), and the Wellcome Trust (Principal Research Fellowship for KJF; Ref: 088130/Z/09/Z).